\documentclass[12pt,twoside]{article}

\pdfoutput=1

\usepackage[usenames]{color}
\usepackage{lscape}
\usepackage{amssymb}
\usepackage{rotating}
\usepackage[T1]{fontenc}
\usepackage{graphicx}
\usepackage{fancyhdr}
\usepackage{amsmath}
\usepackage{ae}
\usepackage{aecompl}
\usepackage{hyperref}

\def\pa{\partial}
\def\nn{\nonumber \\}
\def\ov{\overline}

\def\hgamgam{h\to \gamma\gamma}

\newlength{\dinwidth}
\newlength{\dinmargin}
\begin{document}

\thispagestyle{empty}

\vspace*{1cm}

\centerline{\Large\bf New Regions in the NMSSM with a 125 GeV Higgs }

\vspace*{5mm}

\vspace*{5mm} \noindent
\vskip 0.5cm
\centerline{\bf
Marcin Badziak\footnote[1]{mbadziak@fuw.edu.pl},
Marek Olechowski\footnote[2]{Marek.Olechowski@fuw.edu.pl},
Stefan Pokorski\footnote[3]{Stefan.Pokorski@fuw.edu.pl}
}
\vskip 5mm

\centerline{\em Institute of Theoretical Physics, Faculty of Physics,
University of Warsaw} 
\centerline{\em ul.\ Ho\.za 69, PL--00--681 Warsaw, Poland}

\vskip 1cm

\centerline{\bf Abstract}
\vskip 3mm

It is pointed out that mixing effects in the CP-even scalar sector of the NMSSM can give
6-8 GeV correction to the SM-like Higgs mass in moderate or  large $\tan\beta$ regions with a small value of the singlet-higgs-higgs superfields coupling $\lambda\sim\mathcal{O}(0.1)$. 
This effect comes mainly from the mixing of the SM-like Higgs with lighter singlet.
In the same parameter range, the mixing of the heavy doublet Higgs with the singlet 
may strongly modify the couplings of the singlet-like and the 125 GeV scalars. Firstly, the LEP bounds on a light singlet can be evaded for a large range of its masses. Secondly, the decay rates of both scalars can show a variety of interesting patterns, depending on the lightest scalar mass. In particular, a striking signature of this mechanism can be a light scalar with strongly suppressed (enhanced) branching ratios to $b\bar{b}$ ($gg$, $c\bar{c}$, $\gamma\gamma$) as compared to the SM Higgs with the same mass. The $\gamma\gamma$ decay channel is particularly promising for the search of such a scalar at the LHC.
The 125 GeV scalar can, thus, be accommodated with substantially smaller than in the MSSM radiative
corrections from the stop loops (and consequently, with lighter stops) also for moderate or large $\tan\beta$,
with the mixing effects replacing the standard NMSSM mechanism of increasing the tree level Higgs mass
in the low $\tan\beta$  and large $\lambda$ regime, and with clear experimental signatures of such a mechanism.

\newpage

\section{Introduction}

The discovery of  a SM-like Higgs particle has recently been announced by the LHC experiments \cite{Atlas_discovery,CMS_discovery}.    Although
its properties such as the production   and decay rates into different channels still remain very uncertain \cite{Atlas_moriond, CMS_moriond}, its mass
is established to be around 125 GeV, with only  a couple of GeV uncertainty, and this puts new constraints on the
BSM models.  In the minimal supersymmetric SM (MSSM) the Higgs tree-level quartic coupling  is given by the
electroweak gauge coupling, so that the theory predicts a tree-level upper bound  for the Higgs mass to be equal $M_Z$.
It is well known that loop corrections, mainly from the top-stop loop, can significantly raise the Higgs mass in the MSSM.
The mass of 125 GeV can be accommodated (with loop corrections giving 35 GeV) for certain range of values of the
stop  masses and left-right stop mixing parameter $X_t=A_t-\mu\tan\beta$. That range varies  from $M_{\rm SUSY}\equiv\sqrt{m_{\tilde{t}_1}m_{\tilde{t}_2}}\approx 700$ GeV for the "maximal" stop mixing $X_t\approx\sqrt{6}M_{\rm SUSY}$ to $M_{\rm SUSY}\approx {\mathcal O}(5\, {\rm TeV})$  for $X_t=0$ \cite{softsusy, FeynHiggs} 
\footnote{For a given value of the loop correction, the value of $M_{\rm SUSY}$ depends also on other features of the SUSY spectrum, especially on the gluino mass and the stop mass splitting, the top mass and on unknown higher-order corrections to the Higgs mass. The value of $M_{\rm SUSY}$ is particularly uncertain for $M_{\rm SUSY}\gg1$ TeV, see e.g.~Refs.\cite{GiuStr,ArbeyDjouadi}.  }. 
Such values of the stop mass parameters
are well consistent with the absence so far of any stop signal at the LHC but may look high compared to the standard expectations
based on the naturalness arguments. Awaiting for more experimental progress, one may discard those, after all quite subjective
expectations, or one may hope that a light stop is still hidden in the data, and investigate the ways of reconciling the 125 GeV Higgs
mass with stop mass parameters below the values quoted above.  This necessarily requires a beyond MSSM scheme, with a larger tree-level
Higgs mass than in the MSSM.  (It is useful to note that stop loop radiative corrections $\Delta m_h^{\rm rad}=25\ (30)$ GeV can be reached with $M_{\rm SUSY}\approx
300\ (400)$ GeV for the maximal mixing and with $M_{\rm SUSY}\approx 1.5\ (3)$ TeV for $X_t=0$).  In this context, NMSSM has been discussed in the literature \cite{Gunion,King_bench,Cao,Ellwanger_cNMSSM, PQ_NMSSM125, Stal,Han}.
In the pre-discovery era, NMSSM was discussed mainly as a scenario allowing for a Higgs mass significantly above the values predicted by the MSSM \cite{Ellwanger1,Ellwanger2,Ellwanger3,Ellwanger4,Ellwanger5}.
The attention has been mostly focused on the new tree-level contribution to the Higgs mass coming from the singlet-doublet-doublet coupling
in the superpotential, $\lambda SH_uH_d$, which can be significant for low $\tan\beta$ values and $\mathcal{O}(1)$ values of $\lambda$.  
More recently, already after the discovery of the 125 GeV Higgs, the NMSSM has been discussed in the context of ameliorating the naturalness in the stop sector \cite{NaturalNMSSM_Hall,Li,NaturalNMSSM_King,Gherghetta} also mainly for  the same range of parameters. However, one may think that 125 GeV is close enough to the range expected in the MSSM, so that small corrections to the tree-level mass are worth considering. This is why it is interesting to investigate how significant may be the effect of the singlet-doublet mixing on the Higgs mass in the intermediate and large $\tan\beta$ region, where the MSSM tree-level value is $\sim M_Z$.

In NMSSM, there are three physical neutral CP-even Higgs fields, $H_u$, $H_d$, $S$ which are the real parts of the excitations around the real vevs, $v_u\equiv v \sin\beta$, $v_d\equiv v \cos\beta$, $v_s$ with $v^2=v_u^2 + v_d^2\approx (174 {\rm GeV})^2$,  of the neutral components of doublets $H_u$, $H_d$ and the singlet $S$ (we use the same notation for the doublets and the singlet as for the real parts of their neutral components). It is more convenient for us to work in the  basis $(\hat{h}, \hat{H}, \hat{s})$, where $\hat{h}=H_d\cos\beta + H_u\sin\beta$, $\hat{H}=H_d\sin\beta - H_u\cos\beta$ and $\hat{s}=S$. The $\hat{h}$ field has exactly the same couplings to the gauge bosons and fermions as the SM Higgs field. The field $\hat{H}$ does not couple to the gauge bosons and its couplings to the up and down fermions are the SM Higgs ones rescaled by $\tan\beta$ and $-\cot\beta$, respectively. The mass eigenstates are denoted as $s$, $h$, $H$, with the understanding that $h$ is the SM-like Higgs.

In this paper we point out  that  $\hat{s}-\hat{h}$ mixing  effects can significantly contribute to $m_h$, increasing it by 6-8 GeV, in the intermediate or large
$\tan\beta$ region,  making the NMSSM attractive also in that range of $\tan\beta$. Important for this scenario  is the $\hat{s}-\hat{H}$ mixing, which becomes
significant for larger values of $\tan\beta$ and can lead to a suppression of the $s\rightarrow b\bar{b}$ decay  rate (where $s$ is the singlet-dominated scalar), so that the LEP bounds  on a lighter than 114 GeV scalar coupling to the $Z$ boson based on that decay channel are evaded.
\footnote{The effects of
the $\hat{s}-\hat{h}$ doublet  mixing in the CP-even scalar sector on the SM-like Higgs mass have also been discussed in the literature \cite{PQmixing,Agashe}  
 for low $\tan\beta$ scenario.  The mixing with the heavy doublet is usually ignored as it is small for small $\tan\beta$. Some aspects of the $\hat{s}-\hat{h}$ doublet mixing with large $\tan\beta$ were discussed in Ref.~\cite{Li} but the possibility of a suppression of the $sb\bar{b}$ coupling was not considered there. For an interesting review of non-standard Higgs decays in more general context see Ref.~\cite{review_nonstandardHiggs}.}
The maximum mixing contribution to $m_h$ is then limited by much weaker bounds obtained from the  $s\rightarrow {\rm hadrons}$ signature. The effects discussed in this paper require smallish values of the coupling $\lambda$, $\mathcal{ O}(0.1)$. Thus, the two regions of the
NMSSM parameters, the low $\tan\beta$ one and the one considered here are clearly different.

The $\hat{s}-\hat{H}$ mixing at intermediate and large $\tan\beta$ has another interesting effect. It alters the decay rates of both $s$ and $h$. They become correlated in an interesting
way with  the correction $\Delta m_h^{\rm mix}$  and show a variety of interesting patterns, depending on the lightest scalar mass. In particular, a striking signature of this mechanism can be a light scalar with strongly suppressed (enhanced) branching ratios to $b\bar{b}$ ($gg$, $c\bar{c}$, $\gamma\gamma$) as compared to the SM Higgs with the same mass. The $\gamma\gamma$ decay channel is particularly promising for the search of such a scalar at the LHC.

In section \ref{sec2} we  recall the structure of the  CP-even scalar sector and discuss the effects of the $\hat{s}-\hat{h}$ mixing on the Higgs mass, with the LEP bounds on a light scalar taken into account. In section \ref{sec3} we discuss the potential role of the $\hat{s}-\hat{H}$ mixing and the parameter range for which our mechanism can be relevant. In section \ref{sec4} we give the predictions for the production and decays of the 125 GeV scalar if 5-8 GeV of its mass comes from the $\hat{s}-\hat{h}$ mixing effects. In section \ref{sec:LHCprospects} we briefly discuss the prospects for the discovery of a light scalar at the LHC and in section \ref{sec6} we give a summary of the considered here scenario.

\section{CP-even scalar sector in NMSSM}
\label{sec2}

In this section we recall the necessary for us facts about the CP-even scalar sector of NMSSM \cite{reviewEllwanger}. 
Several versions of NMSSM has been proposed so far \cite{NMSSM_Delgado,GNMSSM,PQ_NMSSM}. We would like to keep our discussion as general as possible so we assume the NMSSM specific part of the superpotential to be:\footnote{Explicit MSSM-like $\mu$-term can also be present in the superpotential but it can always be set to zero by a constant shift of the real component of S.}
\begin{equation}
\label{W_NMSSM}
 W_{\rm NMSSM}= \lambda SH_uH_d + f(S) \,.
\end{equation}
The first term is the source of the effective higgsino mass parameter, $\mu_{\rm eff}\equiv\lambda v_s$ (we drop the subscript ``eff'' in the rest of the paper), while the second term parametrizes various versions of NMSSM. In the simplest version, known as the scale-invariant NMSSM, $f(S)\equiv\kappa S^3/3$.

We assume also quite general pattern of soft SUSY breaking terms (we follow the conventions used in \cite{reviewEllwanger}): 
\begin{equation}
 -\mathcal{L}_{\rm soft}\supset m_{H_u}^2 |H_u|^2 + m_{H_d}^2 |H_d|^2 + m_S^2 |S|^2 + ( A_{\lambda} \lambda H_u H_d S +\frac{1}{3}\kappa A_{\kappa} S^3 + m_3^2 H_u H_d + \frac{1}{2}m_S^{\prime2} S^2 + \xi_S S + {\rm h.c.} )\,. 
\end{equation}
Various versions of NMSSM studied in the literature \cite{NMSSM_Delgado,GNMSSM,PQ_NMSSM} belong to some subclass of the above setup. In the scale-invariant NMSSM, $m_3^2=m_S^{\prime2}=\xi_S=0$.

Let us parametrize the mass matrix of the hatted fields as follows:
\begin{equation}
 \hat{M}^2=
\left(
\begin{array}{ccc}
  \hat{M}^2_{hh} & \hat{M}^2_{hH} & \hat{M}^2_{hs} \\[4pt]
   \hat{M}^2_{hH} & \hat{M}^2_{HH} & \hat{M}^2_{Hs} \\[4pt]
   \hat{M}^2_{hs} & \hat{M}^2_{Hs} & \hat{M}^2_{ss} \\
\end{array}
\right) \,,
\end{equation}
where 
\begin{align}
\label{Mhh}
 &\hat{M}^2_{hh} = M_Z^2\cos^2\left(2\beta\right)+(\delta m_h^2)^{\rm rad} + \lambda^2 v^2\sin^2\left(2\beta\right) \,, \\
\label{MHH}
&\hat{M}^2_{HH} = (M_Z^2-\lambda^2)\sin^2\left(2\beta\right) + \frac{2 B \mu}{\sin\left(2\beta\right)} \,, \\
\label{Mss}
&\hat{M}^2_{ss}=  \frac{1}{2} \lambda v^2 \sin2\beta \left(\frac{\Lambda}{v_s}- \langle\pa^3_S f\rangle\right) + \Upsilon \,,\\
\label{MhH}
& \hat{M}^2_{hH} = \frac{1}{2}(M^2_Z-\lambda^2 v^2)\sin4\beta \,, \\
\label{Mhs}
& \hat{M}^2_{hs} =  \lambda v (2\mu-\Lambda \sin2\beta) \,, \\
\label{MHs}
& \hat{M}^2_{Hs} = \lambda v \Lambda \cos2\beta \,.
\end{align}
Following \cite{ChoiNMSSM}, we introduced $\Lambda\equiv A_{\lambda}+\langle\pa^2_S f\rangle$,
while $B\equiv A_{\lambda}+ \langle\pa_S f\rangle/v_s+m_3^2/(\lambda v_s)$ and $\Upsilon\equiv\langle(\pa^2_S f)^2\rangle + \langle\pa_S f \pa^3_S f\rangle - \frac{\langle\pa_S f \pa^2_S f\rangle}{v_s} + A_{\kappa} \kappa v_s - \frac{\xi_S}{v_s}$.
We neglected all the radiative corrections except those to $\hat{M}^2_{hh}$ which we parametrize by $(\delta m_h^2)^{\rm rad}$.  
The first two terms in eq.~(\ref{Mhh}) are the ''MSSM'' terms, with
\begin{equation}
 (\delta m_h^2)^{\rm rad} \approx \frac{3g^2m_t^4}{8 \pi^2 m_W^2}
\left[\ln\left(\frac{M_{\rm SUSY}^2}{m_t^2}\right)+\frac{X_t^2}{M_{\rm SUSY}^2}
\left(1-\frac{X_t^2}{12M_{\rm SUSY}^2}\right)\right] \,,
\end{equation}
where $M_{\rm SUSY}\equiv\sqrt{m_{\tilde{t}_1}m_{\tilde{t}_2}}$ ($m_{\tilde{t}_i}$
are the eigenvalues of the stop mass matrix  at $M_{\rm SUSY}$ in the $\ov{DR}$ renormalization scheme) and $X_t\equiv
A_t-\mu/\tan\beta$ with $A_t$ being SUSY breaking top trilinear coupling at $M_{\rm SUSY}$.

The third term in eq.~(\ref{Mhh}) is the new tree-level contribution coming from the $\lambda SH_uH_d$ coupling.

We recall that the eigenstates of $\hat{M}^2$ are denoted as $s$, $h$, $H$.
We are interested in the parameter range such that $m_s<m_h<m_H$, so that the $\hat{s}-\hat{h}$ mixing pushes the $m_h$ up. We also require $m_h<2m_s$ to avoid $h\to ss$ decays \cite{Ellwanger_htos}.

Quite generally, the mass of the SM-like Higgs reads:
\begin{equation}
\label{mh2}
 m_h^2=\hat{M}^2_{hh} + (\delta m_h^2)^{\rm mix} \,.
\end{equation}
The $(\delta m_h^2)^{\rm mix}$ term originates mainly from the $\left(\hat{h}, \hat{s}\right)$ mixing and
is positive (negative) when the singlet-dominated scalar is lighter (heavier) then the SM-like Higgs scalar. In the moderate and large $\tan\beta$ regime, the tree-level contribution coming from the $\lambda SH_uH_d$ coupling is suppressed so one has to investigate in detail the potential effects of $(\delta m_h^2)^{\rm mix}$.

\subsection{The effects of the $\hat{s}-\hat{h}$ mixing on the Higgs mass}

In the case of no-mixing with $\hat{H}$, the $\hat{s}-\hat{h}$ mixing is determined by the $2\times2$ block of the mass matrix $\hat{M}^2$:
\begin{equation}
\label{M2hs}
\left(
\begin{array}{cc}
  \hat{M}^2_{hh} & \hat{M}^2_{hs} \\[4pt]
   \hat{M}^2_{hs} & \hat{M}^2_{ss} \\
\end{array}
\right) \,,
\end{equation}
where the entries are given by eqs.~(\ref{Mhh}), (\ref{Mss}) and (\ref{Mhs}). 
The matrix (\ref{M2hs}) is diagonal in the basis $s=\ov{g}_s\hat{h} + \beta_s\hat{s}$, $h=\sqrt{1-\ov{g}_s^2}\hat{h} - \sqrt{1-\beta_s^2}\hat{s}$.

In order to quantify the effect of the $\hat{s}-\hat{h}$ mixing on the Higgs mass it is useful to introduce $\Delta_{\rm mix}$ such that:
\begin{equation}
 m_h=\hat{M}_{hh} + \Delta_{\rm mix} \,.
\end{equation}
Trading $\hat{M}^2_{hh}, \hat{M}^2_{ss}, \hat{M}^2_{hs}$ for the two mass eigenvalues $m_h$, $m_s$ and the coupling $\ov{g}_s$ of the singlet-dominated state to the $Z$ boson (normalized to the corresponding coupling of the SM Higgs), one obtains a simple formula for $\Delta_{\rm mix}$:
\begin{equation}
\label{deltamix}
 \Delta_{\rm mix} = m_h - \sqrt{m_h^2 - \ov{g}^2_s \left(m_h^2-m_s^2\right)} \approx \frac{\ov{g}^2_s}{2} \left(m_h - \frac{m_s^2}{m_h} \right) +\mathcal{O}(\ov{g}^4_s) \,,
\end{equation}
where in the last, approximate equality we used the expansion in $\ov{g}^2_s\ll1$. It is clear from the above formula that a substantial correction to the Higgs mass from the mixing is possible only for not too small couplings of the singlet-like state to the $Z$ boson and that $m_s\ll m_h$ is preferred. However, LEP has provided rather strong constraints on the states with masses below $\mathcal{O}(110)$ GeV that couple to the $Z$ boson because such states could be copiously produced in the process $e^+e^-\to sZ$.

\begin{figure}[t!]
  \begin{center}
    \includegraphics[width=0.49\textwidth]{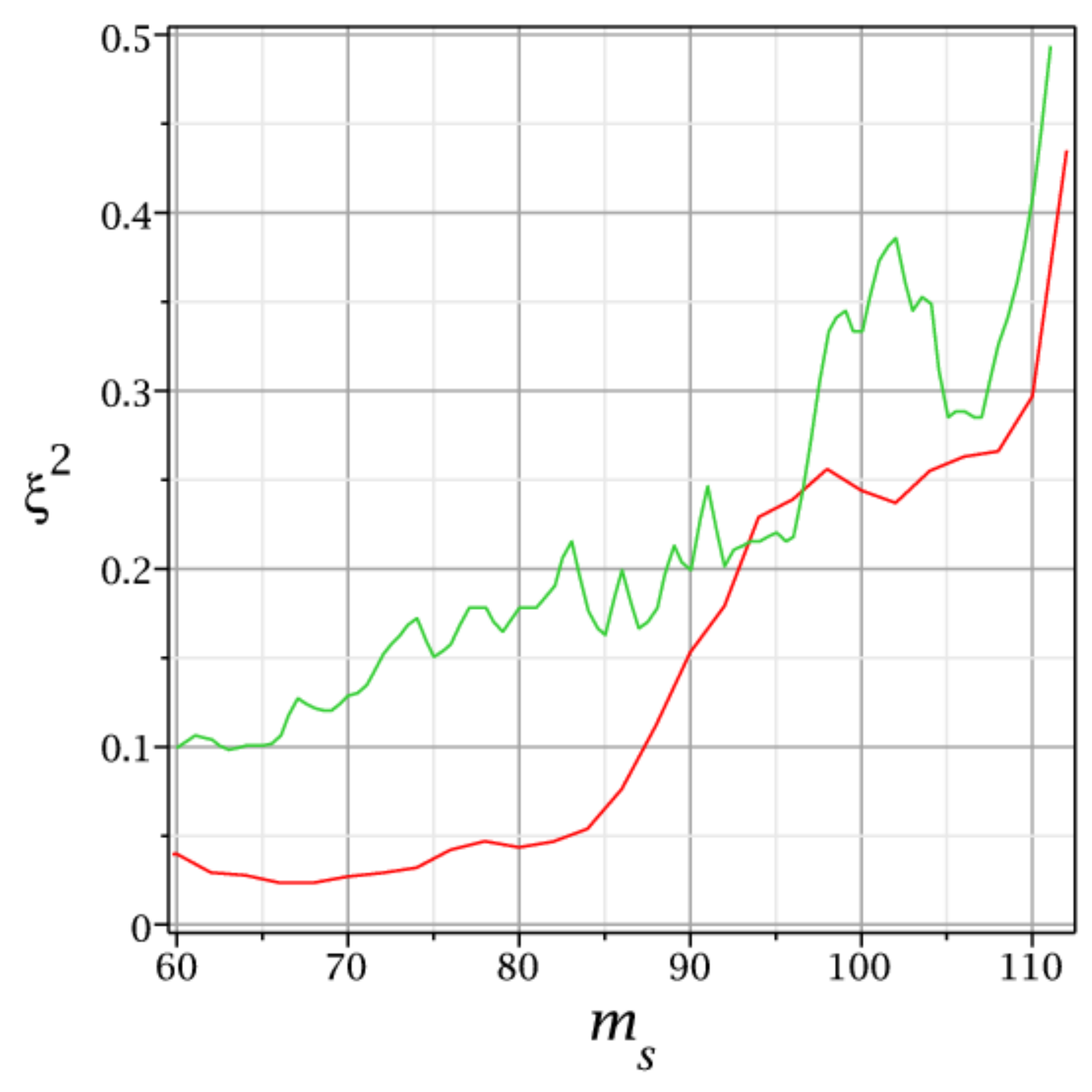}
    \includegraphics[width=0.49\textwidth]{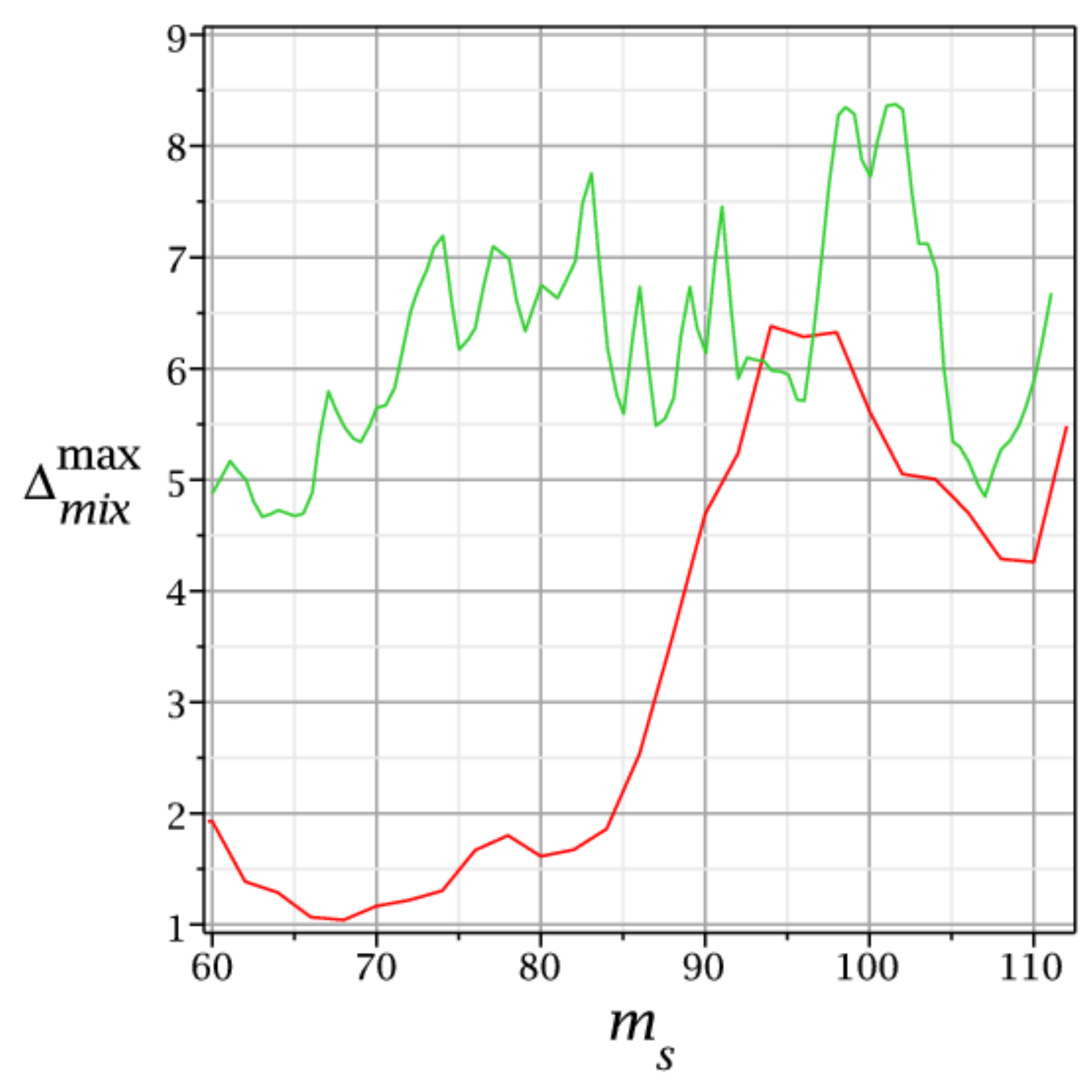}
    \caption{Left: The 95 \% CL upper bounds on $\xi_{b\ov{b}}^2$ (the red line) or $\xi_{jj}^2$ (the green line). 
The red line was obtained using the observed limits presented in the column (a) of Table 14 in Ref.~\cite{LEP_bb}, while the green line corresponds to Figure 2 of \cite{LEP_jj}. Right: The LEP limits translated to the upper limits on $\Delta_{\rm mix}$ using eq.~\ref{deltamix} assuming $\xi_{b\ov{b}}^2=\ov{g}_s^2$ and $\xi_{jj}^2=\ov{g}_s^2$ for the red and green line, respectively. 
    }
    \label{fig:deltamix_ms_LEP}
  \end{center}
\end{figure}

For those LEP searches that rely on the identifications of $b$ and $\tau$ in the final states \cite{LEP_bb}, constraints on $\ov{g}^2_s$ depend on the $s$ branching ratios and the LEP experiments provide constraints on the quantity $\xi^2$ defined as:
\footnote{In the definition \ref{xidef} it is implicitly assumed that the ratio $\Gamma(s\to b\bar{b})/\Gamma(s\to\tau\bar{\tau})$ is the same as for the SM Higgs with the same mass. This is a very good assumption for NMSSM since the $sb\bar{b}$ and $s\tau\bar{\tau}$ couplings (normalized to the corresponding values of the SM Higgs) are the same at tree level.   }
\begin{equation}
\label{xidef}
 \xi_{b\bar{b}}^2\equiv \ov{g}^2_s \times \frac{{\rm BR}(s\to b\bar{b})}{{\rm BR^{\rm SM}}(h\to b\bar{b})}
\end{equation}
The LEP constraints on $\xi_{b\bar{b}}^2$ are reproduced by the red line in the left panel of Figure \ref{fig:deltamix_ms_LEP}.

Since we assume in this subsection that $\hat{s}$ mixes only with $\hat{h}$, all the couplings of $s$ are those of the SM Higgs multiplied by a common factor $\ov{g}_s$. This implies that the branching ratios of $s$ are exactly the same as for the SM Higgs. Therefore, the limits on $\xi_{b\bar{b}}^2$  depicted by the red line  in Figure \ref{fig:deltamix_ms_LEP} are, in fact, also the limits on $\ov{g}_s^2$. Using eq.~(\ref{deltamix}) we can translate the constraints on $\ov{g}_s^2$ into limits for the maximal allowed correction from the mixing, $\Delta_{\rm mix}^{\rm max}$, as a function of $m_s$. These are presented in the right panel of Figure \ref{fig:deltamix_ms_LEP}. Notice that in this case the correction from the singlet-doublet mixing can reach about 6 GeV in a few-GeV interval for $m_s$ around 95 GeV, where the LEP experiments observed the $2\sigma$ excess in the $b\bar{b}$ channel.
 This is interesting since such correction combined with the tree-level values $\sim M_Z$ (for moderate and large $\tan\beta$) gives $m_h\approx 125$ GeV with $\Delta m_h^{\rm rad}\approx 30$ GeV.
However, for $m_s\lesssim90$ GeV the allowed value of $\Delta_{\rm mix}^{\rm max}$ drops down very rapidly to very small values.

\begin{figure}[t!]
  \begin{center}
    \includegraphics[width=0.49\textwidth]{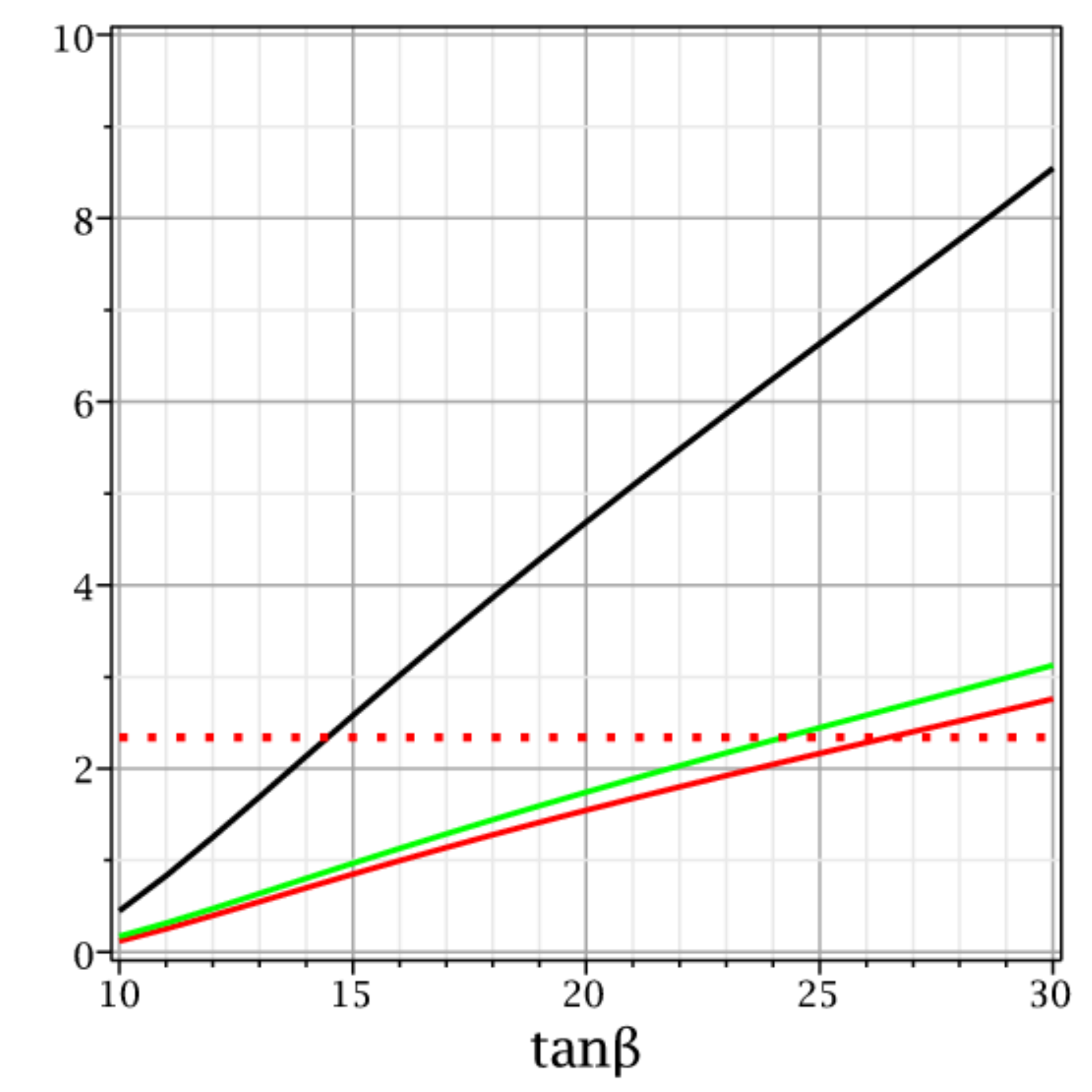}
    \caption{ $\Delta_{\rm mix}$ [GeV] (black solid line), $\ov{g}_s^2\times10$ (green solid line) and $\xi_{b\bar{b}}^2\times10$ (red solid line) as a function of $\tan\beta$. The remaining parameters are fixed to be: $m_h=125$ GeV and $\mu=150$ GeV, $m_s=95$ GeV, $m_H=1500$ GeV, $\Lambda=1200$ GeV and $\lambda=0.08$. The constraint on $\xi_{b\bar{b}}^2\times10$ is depicted by the dotted red line. The regions with the solid red line below the dotted red line are allowed by the LEP data. 
    }
    \label{fig:deltamix_tanb_hsonly}
  \end{center}
\end{figure}

In Figure \ref{fig:deltamix_tanb_hsonly} we present an example of the NMSSM parameters for which mixing with $\hat{H}$ is negligible and $\Delta_{\rm mix}\approx6$ GeV can be obtained. Note that the $\hat{h}-\hat{s}$ mixing, thus also $\Delta_{\rm mix}$, grows with $\tan\beta$ as a consequence of the suppression of the second term in the parenthesis in $\hat{M}^2_{hs}$ at large $\tan\beta$, see eq.~(\ref{Mhs}). This example demonstrates also the fact that $\lambda$ is generically at most $\mathcal{O}(0.1)$. Larger values of $\lambda$ typically lead to too large $\hat{M}^2_{hs}$  (after taking into account the LEP limit on the chargino mass which imply $\mu\gtrsim100$ GeV) leading to a negative determinant of the mass matrix. Therefore, this scenario is the most natural at moderate and large $\tan\beta$. 
\footnote{In the scenario with $\Delta_{\rm mix}>0$, values of $\lambda\sim0.6$ that lead to substantial tree-level contribution to $m_h$ at small $\tan\beta$ can only be obtained if $(2\mu-\Lambda\sin(2\beta))$ (which enters $\hat{M}^2_{hs}$) is finely-tuned to be below ${\mathcal O}(10 {\rm GeV})$.}

It is also clear from Fig.~\ref{fig:deltamix_ms_LEP}  that similar correction  $\mathcal{O}(5)$ GeV to the Higgs mass can be obtained from the 
$\hat{s}-\hat{h}$ mixing  for a larger range of the singlet-dominated  scalar mass $m_s$,  provided one can evade
the LEP bounds given by the red curve in the left panel of Fig. \ref{fig:deltamix_ms_LEP} by suppressing the $sb\bar {b}$ and  $s\tau\bar{\tau}$ couplings. This is because in such a case, $s$ decays predominantly into charm quarks and gluons and $b$-tagging cannot be used to enhance the signal over background ratio so the most stringent constraints on $\ov{g}_s^2$ come from the flavour independent Higgs searches in hadronic final states at LEP \cite{LEP_jj}. Those searches give constraints on a quantity $\xi_{jj}^2$ defined as:
\begin{equation}
 \xi_{jj}^2\equiv \ov{g}^2_s \times {\rm BR}(s \to jj) \,,
\end{equation}
which are reproduced by the green line in the left panel of Fig.~\ref{fig:deltamix_ms_LEP}. Noting that for suppressed $sb\bar {b}$ and  $s\tau\bar{\tau}$ couplings, ${\rm BR}(s \to jj)\approx1$ so $\xi_{jj}^2\approx \ov{g}^2_s$, we can translate those constraints into the upper bound on $\Delta_{\rm mix}$.  
Indeed, the upper bound $\Delta^{\rm max}_{\rm mix}$ is then given by the green
curve  in the right panel of Fig.\ \ref{fig:deltamix_ms_LEP}.

We show in the next section that  $\hat{s}-\hat{H}$
mixing can significantly change the decay rates of $s$ and also of $h$. 
\footnote{
A suppression of the $s\to b\bar{b}$ decay rate is possible for any value of $m_s$ but for $m_s$ in the few-GeV interval around 95 GeV there is no gain in $\Delta^{\rm max}_{\rm mix}$ because the red and green curves in Fig.\ \ref{fig:deltamix_ms_LEP} practically overlap there.
} 
Firstly, the $\Delta^{\rm max}_{\rm mix}$ shown by the green line in  Fig.\ \ref{fig:deltamix_ms_LEP}  can then be obtained for a broad range $60\ {\rm GeV}<m_s<110\ {\rm GeV}$, and secondly the decay rates of $s$ and $h$ can have interesting patterns.

\section{Singlet mixing with both doublets and the suppression of the $sb\bar{b}$ coupling}
\label{sec3}

We now go back to the general case in which mixing with $\hat{H}$ may be present. As in the previous section, we begin with the implications following from the general structure of the mass matrix. Mixing with $\hat{H}$ leads to the modification of Higgs couplings to fermions.  Denoting the mass-eigenstates $s$, $h$, $H$ by $x=\ov{g}_x\hat{h} + \beta^{(H)}_x\hat{H} + \beta^{(s)}_x\hat{s}$ we get
\begin{align}
\label{Cb}
 &C_{b_x}=\ov{g}_x+\beta^{(H)}_x\tan\beta \,, \\
 &C_{t_x}=\ov{g}_x-\beta^{(H)}_x\cot\beta \,, \\
 &C_{V_x}=\ov{g}_x \,,
\end{align}
where $x$ is $s$, $h$ or $H$. Note that the couplings to the vector bosons depend only on the $\hat{h}$ components, as in the case of only $(\hat{h}, \hat{s})$ mixing discussed in the previous subsection.

In the region of moderate and  large $\tan\beta$  even small component of $\hat{H}$ in the singlet-dominated Higgs may give a large contribution to the couplings to $b$ quark due to $\tan\beta$ enhancement. On the other hand, the couplings to the up-type quarks are almost the same as those to the gauge bosons, $C_{t_x} \approx C_{V_x}$. Particularly interesting is the case when $\ov{g}_s$ has the opposite sign to $\beta^{(H)}_s$ because then $C_{b_s}\ll C_{t_s}, C_{V_s}$ is possible. In the regime $C_{b_s}\ll C_{t_s}, C_{V_s}$, the (otherwise dominating) $s$ branching ratios to $b\bar{b}$ and $\tau\bar{\tau}$ are strongly suppressed and $s$ decays mainly to $gg$ and $c\bar{c}$. The ratio $\Gamma(s\to gg)/\Gamma(s \to c\bar{c})$ is roughly the same as for the SM Higgs so e.g. for $m_s=90$ GeV it equals about $1.5$ \cite{xsecWG,hdecay} and approximately scales like $m_s^2$ for other masses \cite{DjouadiSMreview}. In this regime the standard LEP Higgs searches \cite{LEP_bb} that used $b$-tagging cannot be applied to constrain this scenario. In such a case, the most stringent constraints comes from the flavour independent search for a Higgs decaying into two jets at LEP \cite{LEP_jj}. These constraints are weaker and allow for values of $\ov{g}_s^2$ above 0.3 for $m_s$ around 100 GeV and the limit rather slowly improves as $m_s$ goes down, as seen from the left panel of Figure \ref{fig:deltamix_ms_LEP}. 
In consequence, the constraints on $\Delta_{\rm mix}^{\rm max}$ are also weaker. As can be seen from the right panel of Figure \ref{fig:deltamix_ms_LEP}, when $s\to b\bar{b}$ decays are suppressed $\Delta_{\rm mix}$ above 5 GeV is viable for a large range of $m_s$ with a maximum of about 8 GeV for $m_s$ around 100 GeV.

We should also comment on the fact that for $C_{b_s}\ll C_{t_s}, C_{V_s}$ the $s$ branching ratios to the gauge bosons are also enhanced (with respect to the SM Higgs predictions) by a factor that can exceed 10. In spite of such a large enhancement the Higgs searches in these channels performed at LEP \cite{LEP_gamma} are less constraining than the above-discussed searches with hadronic decays. On the other hand, the LHC searches in the diphoton channel may, in principle, have a potential to give additional constraints on this scenario (i.e. reduce the allowed value of $\Delta_{\rm mix}^{\rm max}$). In fact, $s\to\gamma\gamma$ decays could already be seen at the LHC but the SM Higgs searches in the diphoton channel have been performed only for masses above 110 GeV. This will be discussed in more detail in section \ref{sec:LHCprospects}

In the above discussion we assumed that the strong suppression of the $s$ coupling to $b$ quarks is possible. Let us now discuss in which part of the NMSSM parameter space such situation may hold. As already stated, this may happen only for not too small values of $\tan\beta$ and a negative ratio $\beta^{(H)}_s/\ov{g}_s$. This ratio can be expressed in terms of the $\hat{M}^2$ entries and $m_s$ in the following way:
\begin{equation}
 \frac{\beta^{(H)}_s}{\ov{g}_s}=\frac{ \hat{M}^2_{Hs} \left(\hat{M}^2_{hh} - m_s^2\right) - \hat{M}^2_{hs}\hat{M}^2_{hH}   }{  \hat{M}^2_{hs} \left(\hat{M}^2_{HH} - m_s^2\right) - \hat{M}^2_{Hs}\hat{M}^2_{hH}  } \,.
\end{equation}
At large $\tan\beta$, $\hat{M}^2_{hH}\approx-2(M^2_Z-\lambda^2 v^2)/\tan\beta$ is very small
 so the second terms in the numerator and the denominator are typically subdominant 
\footnote{Strictly speaking, the second terms in the numerator and the denominator can dominate for $\Lambda\to0$ because then $\hat{M}^2_{Hs}\to0$. In such a case $\beta^{(H)}_s/\ov{g}_s$ is negative if $\hat{M}^2_{hH}>0$ which is possible only if $\lambda^2 v^2> M_Z^2$. However, for $\lambda^2 v^2> M_Z^2$ and $\Lambda\to0$ the mass matrix has a negative eigenvalue if $\hat{M}^2_{ss}<\hat{M}^2_{hh}$ (which is a necessary condition for $s$ to be lighter than $h$).}
which means that $\beta^{(H)}_s/\ov{g}_s$ is negative if $\hat{M}^2_{Hs} \hat{M}^2_{hs} <0$ (we recall that $m_s^2<\hat{M}^2_{hh}$ in our case) which leads to the following condition for the NMSSM parameters:
\begin{equation}
 \Lambda(\mu\tan\beta-\Lambda)\gtrsim0 \,,
\end{equation}
which is satisfied only if $\mu\Lambda>0$.
In the following discussion we will assume, without loss of generality, $\Lambda>0$ and $\mu>0$.
It is straightforward to show  in the limit of large $\tan\beta$ that $C_{b_s}$ may vanish only if 
\begin{equation}
\label{cb0cond}
 r^2>\frac{2\Lambda^2}{\mu^2 } \,,
\end{equation}
where
\begin{equation}
  r^2\equiv\frac{\hat{M}^2_{HH} - m_s^2}{\hat{M}^2_{hh} - m_s^2} \,.
\end{equation}
If the condition (\ref{cb0cond}) is satisfied then $C_{b_s}\approx0$ corresponds to two values of $\tan\beta$:
\begin{equation}
 \tan\beta\approx\frac{\mu r^2}{\Lambda }\left( 1\pm\sqrt{1-\frac{2\Lambda^2}{\mu^2 r^2} }\right) \,,
\end{equation}
which in the limit $r^2\gg\frac{2\Lambda^2}{\mu^2 }$ are given by:
\begin{equation}
 \tan\beta\approx\frac{\Lambda}{\mu}\left(1+\frac{\Lambda^2}{2\mu^2r^2}\right) \qquad \vee \qquad \tan\beta\approx\frac{2\mu r^2}{\Lambda} \,.
\end{equation}

\begin{figure}[t!]
  \begin{center}
    \includegraphics[width=0.49\textwidth]{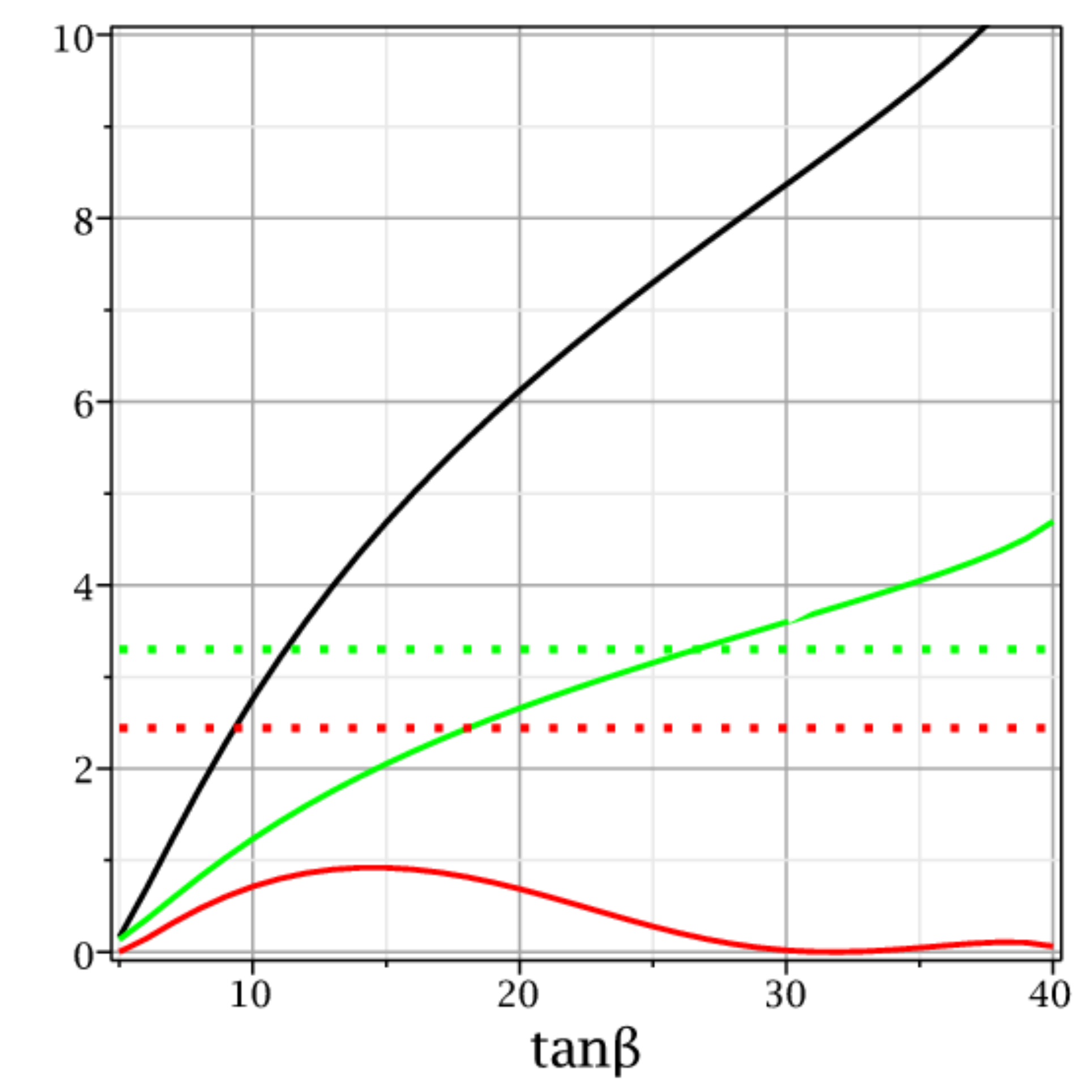}
    \includegraphics[width=0.49\textwidth]{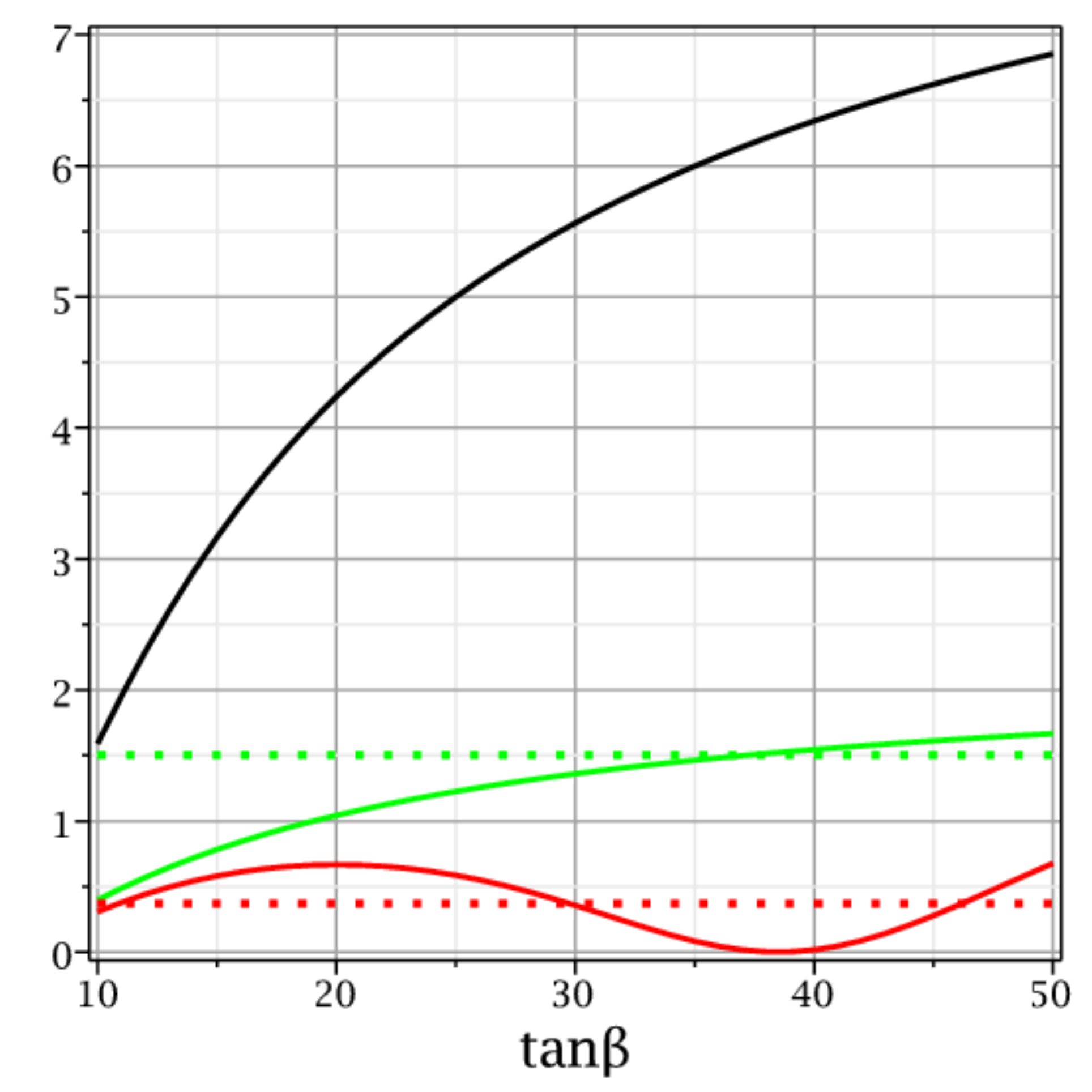}
    \caption{ $\Delta_{\rm mix}$ [GeV] (black solid line), $\ov{g}_s^2\times10$ (green solid line) and $\xi_{b\bar{b}}^2\times10$ (red solid line) as a function of $\tan\beta$. For easy reference the LEP constraints are depicted by the dotted lines with the same color coding as in Figure \ref{fig:deltamix_ms_LEP}. A point in parameter space is consistent with the LEP data if the red solid line is below the red dotted line and the green solid line is below the green dotted line. In both panels $m_h=125$ GeV and $\mu=150$ GeV. In the left panel: $m_s=100$ GeV, $m_H=500$ GeV, $\Lambda=600$ GeV and $\lambda=0.06$. In the right panel: $m_s=75$ GeV, $m_H=1000$ GeV, $\Lambda=800$ GeV and $\lambda=0.08$.
    }
    \label{fig:deltamix_tanb_Heffects}
  \end{center}
\end{figure}

Let us now demonstrate some numerical examples in which the suppression of $C_{b_s}$ is present and substantial values of $\Delta_{\rm mix}$ is obtained without violating the LEP constraints.
In Figure \ref{fig:deltamix_tanb_Heffects} a $\tan\beta$-dependence of $\Delta_{\rm mix}$ is presented. It is clear from this Figure that substantially larger $\Delta_{\rm mix}$ is consistent with the LEP data due to the suppression of the $sb\bar{b}$ coupling. In the left panel, $m_s=100$ GeV and $\Delta_{\rm mix}$ can be almost 8 GeV. The role of the suppression of the $sb\bar{b}$ coupling is even more important for lighter singlet-dominated states. In the right panel, $m_s=75$ GeV and  $\Delta_{\rm mix}$ can reach 6 GeV at large $\tan\beta$, while without the suppression it would be below 2 GeV.

It can also be seen from Figure \ref{fig:deltamix_tanb_Heffects} that there exist values of $\tan\beta$ for which the $sb\bar{b}$ coupling is strictly zero. Nevertheless, such a strong suppression is not necessary to avoid the LEP constraints. In fact it is enough to suppress BR($s\to b\bar{b}$) by about 25\% in the case of $m_s=100$ GeV and by a factor of three for $m_s=75$ GeV. This implies rather large range of $\tan\beta$  with significant correction from the mixing consistent with the LEP data.

In our analysis we use the eigenvalues of the Higgs mass matrix as input parameters while the diagonal entries of this matrix are output parameters. Such procedure is justified because any values of the diagonal entries can be obtained by adjusting the soft terms in appropriate way. However, it is natural to ask whether the required values of soft terms are reasonable. One cannot answer this question in a model-independent way so let us focus on the no-scale version of NMSSM which is the most popular one and calculate the soft terms in some representative examples. In such a case, $\mu=\lambda v_s$, $B=A_{\lambda}+\kappa v_s$ and $\Lambda=A_{\lambda}+2\kappa v_s$. Requiring  the correct electroweak minimum,  for the parameters used in the left panel  of Figure \ref{fig:deltamix_tanb_Heffects} one obtains for $\tan\beta=25$ (corresponding to $\Delta_{\rm mix}\approx7.3$ GeV):
\begin{align}
 &A_{\kappa}=-2111\ {\rm GeV}, \quad A_{\lambda}=-467\ {\rm GeV}, \quad \kappa=0.213, \nn
 &m_{H_u}^2=-(162\ {\rm GeV})^2, \quad m_{H_d}^2=(480\ {\rm GeV})^2, \quad m_{S}^2=(109\ {\rm GeV})^2  \,,
\end{align}
while for the parameters used in the right panel of Figure \ref{fig:deltamix_tanb_Heffects} one obtains for $\tan\beta=35$ (corresponding to $\Delta_{\rm mix}\approx6$ GeV):   
\begin{align}
 &A_{\kappa}=-2427\ {\rm GeV}, \quad A_{\lambda}=-419\ {\rm GeV}, \quad \kappa=0.325, \nn
 &m_{H_u}^2=-(161\ {\rm GeV})^2, \quad m_{H_d}^2=(990\ {\rm GeV})^2, \quad m_{S}^2=(85\ {\rm GeV})^2  \,.
\end{align}
From the above two examples it should be clear that large values of $\Delta_{\rm mix}$ can be obtained for rather natural values of the soft parameters. In particular, the values of $\kappa$ can be consistent with the upper bound, $\kappa_{\rm max}\approx0.65$ (for $\lambda\lesssim0.1$) \cite{reviewEllwanger}, from the requirement of perturbativity up to the GUT scale.

\section{Production and decays of the 125 GeV Higgs}
\label{sec4}

The mixing effects affect not only the branching ratios and production cross-section of $s$ but also those of $h$. Moreover, they are correlated so the scenario may be tested also by the measurements of the signal strengths for the 125 GeV Higgs. In order to set a notation let us define the signal strengths modifiers as:
\begin{equation}
 R_i^{(h)}\equiv\frac{\sigma(pp\to h)\times {\rm BR}(h \to i)}{\sigma^{\rm SM}(pp\to h)\times {\rm BR}^{\rm SM}(h \to i)} \,.
\end{equation}

In the case of the $\hat{h}-\hat{s}$ mixing, with the effects of $\hat{H}$ neglected, all the $h$ couplings are multiplied by a common factor $\sqrt{1-\ov{g}_s^2}$. This implies that all the $h$ branching ratios are the same as for the SM Higgs while the production cross-section (in all channels) is smaller by a factor $1-\ov{g}_s^2$ so $R_i^{(h)}=1-\ov{g}_s^2$ for all channels. This means that, after taking into account the LEP constraints, $\Delta_{\rm mix}>5$ GeV implies $0.75\lesssim R_i^{(h)}\lesssim 0.83$.   

In the full $3\times 3$ mixing case at large $\tan\beta$, the couplings to the up-type quarks are almost the same as those to the gauge bosons, $C_{t_h}\approx C_{V_h}=\sqrt{1-\ov{g}_s^2}$ so the production cross-section is still smaller than the SM prediction by a factor $1-\ov{g}_s^2$. However, the couplings to the down-type fermions can be substantially modified, as seen from eq.~(\ref{Cb}). Since the $b\bar{b}$ channel dominates the decays of the 125 GeV SM Higgs (${\rm BR^{\rm SM}}(h\to b\bar{b})\approx58\%$ and ${\rm BR^{\rm SM}}(h\to \tau\bar{\tau})\approx6\%$) such modifications lead to important effects for all the other branching ratios. If $\beta^{(H)}_h/\ov{g}_h$ is negative (positive) then the $h$ couplings to $b$ and $\tau$ are smaller (larger) than in the SM which leads to the enhancement (suppression) of the Higgs branching ratios to the gauge bosons and two photons. This ratio is given by
\begin{equation}
 \frac{ \beta^{(H)}_h }{ \ov{g}_h } = - \frac{ \hat{M}_{hH}^2 \left(m_h^2 - \hat{M}_{ss}^2\right) +  \hat{M}_{hs}^2 \hat{M}_{Hs}^2  }{  \left( \hat{M}_{HH}^2 - m_h^2\right) \left( m_h^2 - \hat{M}_{ss}^2 \right) + \left( \hat{M}_{Hs}^2 \right)^2 } 
\end{equation}
and its sign is:
\begin{equation}
 {\rm sgn} \left( \frac{ \beta^{(H)}_h }{ \ov{g}_h } \right) = - {\rm sgn} \left( \hat{M}_{hH}^2 + \frac{ \hat{M}_{hs}^2 \hat{M}_{Hs}^2 }{\left( m_h^2 - \hat{M}_{ss}^2 \right) } \right) \,.
\end{equation}
Since $\hat{M}_{hH}^2$ is small, the enhancement (suppression) of the $h$ coupling to $b$ requires $\hat{M}^2_{Hs} \hat{M}^2_{hs} <0$ ($>0$). Note that this is the opposite condition to that for the $s$ coupling so if the $s$ coupling to $b$ is enhanced  (suppressed) then the $h$ coupling to $b$ is suppressed (enhanced). As it was discussed in the previous section, for the $m_s$ in the range between about 90 and 105 GeV, $\Delta_{\rm mix}$ can exceed 5 GeV with the LEP constraints satisfied independently of the $sb\bar{b}$ coupling 
and both discussed above options are interesting. 
From the current experimental viewpoint the suppressed $h$ coupling to $b$ is more welcome in order to compensate the suppression of the $h$ production cross-section and end up with $R_{VV}^{(h)}\approx1$ (where $V=W$ or $Z$).
\footnote{Such a scenario was investigated for low $\tan\beta$ in \cite{Ellwanger_Rgam_h} with a special attention to possible $\gamma\gamma$ rate enhancement for the 125 GeV Higgs.}
  However, given the present tension between the CMS and ATLAS results the case with the enhanced $h$ coupling to $b$ is certainly not excluded.    

The predictions for $R_{\gamma\gamma}^{(h)}$ are very similar to $R_{VV}^{(h)}$ because the reduced couplings to top and $W$ (which contribute to the $\hgamgam$ decay in the SM) are almost the same at large $\tan\beta$, $C_{t_h}\approx C_{V_h}$. The enhancement of $R_{\gamma\gamma}^{(h)}$ over $R_{VV}^{(h)}$, which is preferred by the ATLAS data, is possible only if contributions of SUSY particles to the $\hgamgam$ decay width is non-negligible. It was shown in Refs.~\cite{diphoton_chargino,ChoiNMSSM} that such enhancement can be substantial for light higgsinos and $\lambda\sim\mathcal{O}(1)$. However, we found that this effect is small in our case since $\lambda$ is required to be $\mathcal{O}(0.1)$ at most. The most promising way to obtain the $\gamma\gamma$ enhancement would be the presence of very light staus with strong left-right mixing which may be possible if $\tan\beta$ is large \cite{CarenaStau}.

\subsection{$s$ with strongly suppressed couplings to $b$ and $\tau$}

It is crucial to note that the couplings of $h$ and $s$ to $b$ are correlated. It is the purpose of this subsection to investigate the implications of the strongly suppressed $s$ couplings to $b$ for the production rates of $h$.

\begin{figure}[t!]
  \begin{center}
    \includegraphics[width=0.49\textwidth]{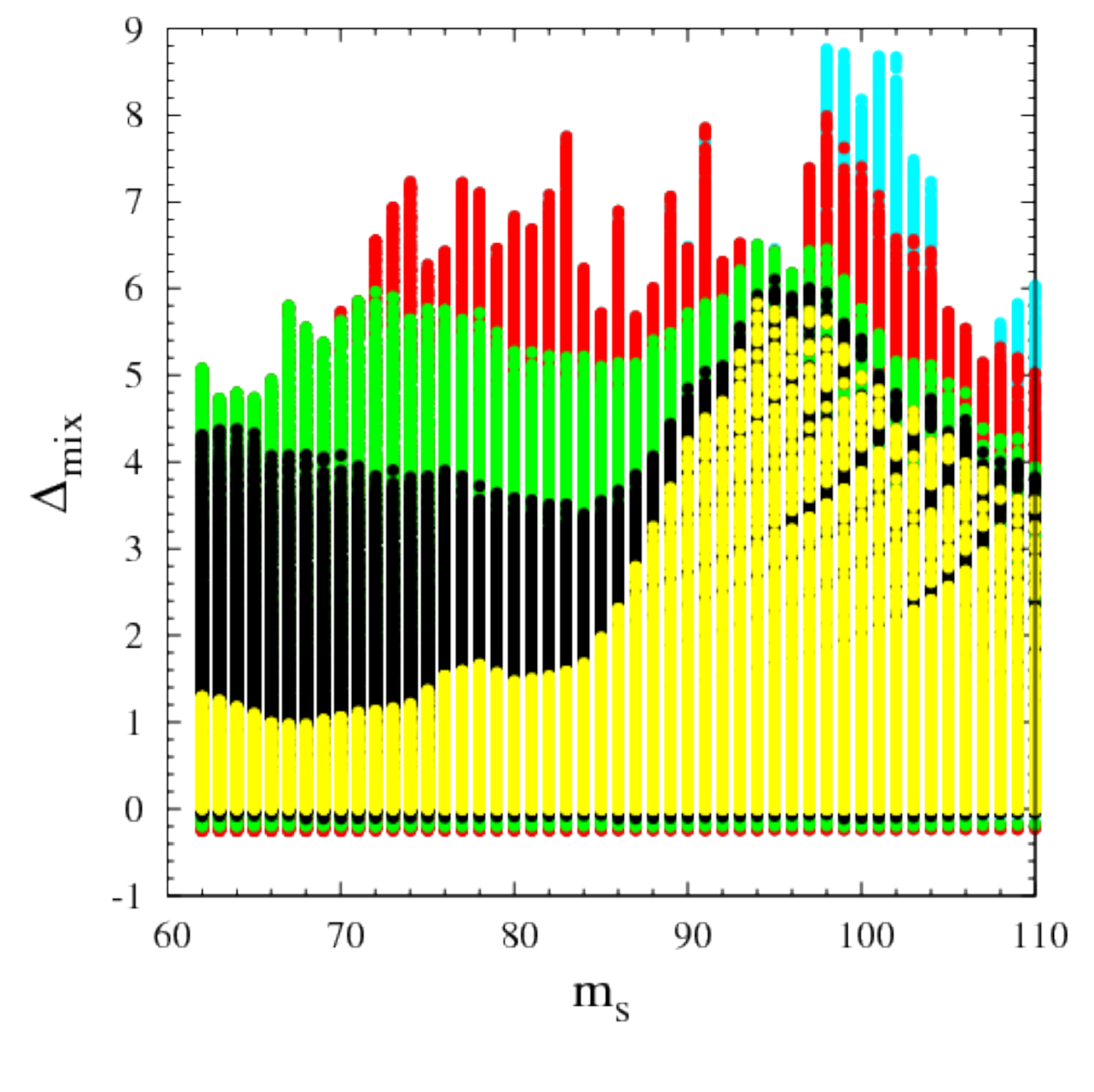}
   
    \caption{Results of the numerical scan presented in the $\Delta_{\rm mix}$-$m_s$ plane. Different colours correspond to different values of $R_{VV}^{(h)}$. The blue points are characterised by $R_{VV}^{(h)}<0.5$ while for the red, green, black  and yellow points $R_{VV}^{(h)}$ is larger than 0.5, 0.7, 0.8 and 1, respectively.  
              The points with larger values of $R_{VV}^{(h)}$ are overlaid on the points with smaller $R_{VV}^{(h)}$. All the points satisfy the LEP constraints.
    }
    \label{fig:deltamix_ms_scan}
  \end{center}
\end{figure}

In order to study quantitatively the correlation between the correction to the Higgs mass from mixing and the production rates for $h$ we performed a numerical scan over the NMSSM parameter space for various values of $m_s$ and $m_H$ while keeping fixed $m_h=125$ GeV. In the scan we also fixed $\mu=150$ GeV. For other values of $\mu$ the results of the scan are the same provided that the following transformation of parameters is used:
\begin{equation}
 \mu \to k\mu\,, \qquad \lambda \to \lambda/k\,, \qquad \Lambda \to k\Lambda \,.
\end{equation}
 This is because $\hat{M}_{hs}^2$ and $\hat{M}_{Hs}^2$ are invariant under the above transformation while $\hat{M}_{hH}^2$ is only marginally affected so its impact on the numerical results is negligible.
The remaining parameters where scanned on a grid, see Table \ref{tab:scan} for the scanned parameters ranges and step sizes. In order to emphasize that obtaining substantial values of $\Delta_{\rm mix}$ does not require any fine-tuning the grid is not dense, as clearly seen from Table \ref{tab:scan}. In Figure \ref{fig:deltamix_ms_scan}  a scatter plot of $\Delta_{\rm mix}$ versus $m_s$ is presented. The LEP constraints discussed before have been taken into account. It can be seen that $\Delta_{\rm mix}$ up to about 9 GeV can be obtained for $m_s\approx100$ GeV but for such large values of $\Delta_{\rm mix}$ $R_{VV}^{(h)}<0.5$ is predicted,
which is in tension with the LHC Higgs data.\footnote{Notice that maximal values of $\Delta_{\rm mix}$ for a given $m_s$ in Figure \ref{fig:deltamix_ms_scan} are slightly larger than the corresponding values in the right panel of Figure \ref{fig:deltamix_ms_LEP}. This is because in Figure \ref{fig:deltamix_ms_LEP} BR$(s\to jj)=1$, i.e.\ $\ov{g}_s^2=\xi_{jj}^2$, is assumed, while for the points from the numerical scan that are consistent with the LEP data the $sb\bar{b}$ and $s\tau\bar{\tau}$ couplings are not exactly zero so BR$(s\to \tau\bar{\tau})>0$ leading to $\ov{g}_s^2>\xi_{jj}^2$.  } 
Nevertheless, demanding $R_{VV}^{(h)}>0.5$, $\Delta_{\rm mix}$ about 8 GeV can be reached. Notice also that $\Delta_{\rm mix}\gtrsim5$ GeV with $R_{VV}^{(h)}>0.7$, which is well consistent with the LHC data within the experimental errors, can be obtained for a wide range of values between $m_h/2$ and 105 GeV.

\begin{table}[t!]
\centering
 \begin{tabular}{|c|c|c|c|c|}
\hline
     & $m_H$ [GeV] & $\lambda$ & $\Lambda$ [GeV] & $\tan\beta$ \\ \hline
Minimal value  & 250 & 0.05 & 100 & 10 \\ \hline
Maximal value  & 2000 & 0.15 & 3000 & 60 \\ \hline
Step size  & 250 & 0.01 & 100 & 5 \\ \hline
 \end{tabular} 
\caption{The parameter ranges and step sizes used in the numerical scan.}
\label{tab:scan}
\end{table}

The reduction of $R_{VV}^{(h)}$ is due to the  $h$ production cross-section suppressed by a factor $1-\ov{g}_s^2$ and the suppressed BR($h\to VV$), as a consequence of the enhanced $hb\bar{b}$ coupling. However, for $m_s$ between about 90 and 105 GeV, where the LEP constraints on $\ov{g}_s^2$ are not so strong and $m_s$ is still significantly below 125 GeV, suppression of the $sb\bar{b}$ coupling is not necessary for obtaining substantial values of $\Delta_{\rm mix}$. Therefore, in that range $R_{VV}^{(h)}>1$ can be obtained with $\Delta_{\rm mix}\gtrsim5$ GeV. Such solutions are characterised by the enhanced $sb\bar{b}$ coupling and suppressed $hb\bar{b}$ coupling.  

Since for moderate and large $\tan\beta$, $C_V\approx C_t$, the predictions for  $R_{\gamma\gamma}^{(h)}$ is almost the same as for $R_{VV}^{(h)}$, presented in Figure \ref{fig:deltamix_ms_scan}. After taking into account the higgsino contribution to the $\hgamgam$ decay rate $R_{\gamma\gamma}^{(h)}$ becomes slightly larger than $R_{VV}^{(h)}$. For $\mu=150$ GeV, which was used in the numerical scan, $R_{\gamma\gamma}^{(h)}$ is typically enhanced with respect to $R_{ZZ}^{(h)}$ by a few percent. 

We should stress that our analysis is performed at tree level. It is well known that at large $\tan\beta$ SUSY threshold correction to the bottom quark Yukawa coupling may be substantial \cite{Hall, Carena}. If those corrections act in such a way that the loop-corrected $C_{b_h}$ is smaller than the tree-level value then the $h$ branching ratio into gauge bosons is enhanced. Thus, in principle some regions of the NMSSM parameter space may exist in which $\Delta_{\rm mix}$ reaches 8 GeV and $R_{VV}^{(h)}$ is around one, without violation of the LEP constraints. However, a detailed study of such corrections is beyond the scope of this paper.

\section{Prospects for discovery of $s$ at the LHC}
\label{sec:LHCprospects}

Let us now discuss prospects for discovery of $s$ at the LHC. What the experiments observe is the product of the production cross-section and the branching ratios:
\begin{equation}
 R_i^{(s)}\equiv\frac{\sigma(pp\to s)\times {\rm BR}(s \to i)}{\sigma^{\rm SM}(pp\to h)\times {\rm BR}^{\rm SM}(h \to i)}
\end{equation}
If the $s$ branching ratio to $b\bar{b}$ is not strongly modified as compared to that of the SM Higgs, the signal strengths in all channels are universally suppressed $R_i^{(s)}\approx \ov{g}_s^2$. For $m_s$ in the range 90-105 GeV, where $\Delta_{\rm mix}\gtrsim5$ GeV is possible without strong $sb\bar{b}$ coupling suppression, $R_i^{(s)}\lesssim0.25$.
\footnote{Particularly interesting possibility is the singlet-like Higgs with mass about 98 GeV because it can explain the LEP excess in the $b\bar{b}$ channel \cite{Higgs_LEP_LHC}.}
In that range of $m_s$ the LHC experiments have the best sensitivity in the $s\to b\bar{b}$ decay channel (the $\gamma\gamma$ and $\tau\bar{\tau}$ channels may also be relevant, especially for $m_s\gtrsim100$ GeV). The LHC experiments do not provide limits, nor the expected sensitivities, for the masses below $110$ GeV in their searches for the SM Higgs. However, the sensitivity of the search in the $b\ov{b}$ channel very weakly depends on $m_s$ in this range so one can estimate that the expected sensitivity to $R_{b\bar{b}}^{(s)}$ is about 0.9 with the data that have been analysed so far i.e. 5 fb$^{-1}$ of the 7 TeV data and 13 fb$^{-1}$ of the 8 TeV data \cite{ATLAS_bb, CMS_bb}. From a naive extrapolation to higher luminosities one expects that about 200 fb$^{-1}$ of the 14 TeV run will be required to test this scenario. 

\subsection{$s$ with strongly suppressed couplings to $b$ and $\tau$}

\begin{figure}[t!]
  \begin{center}
    \includegraphics[width=0.49\textwidth]{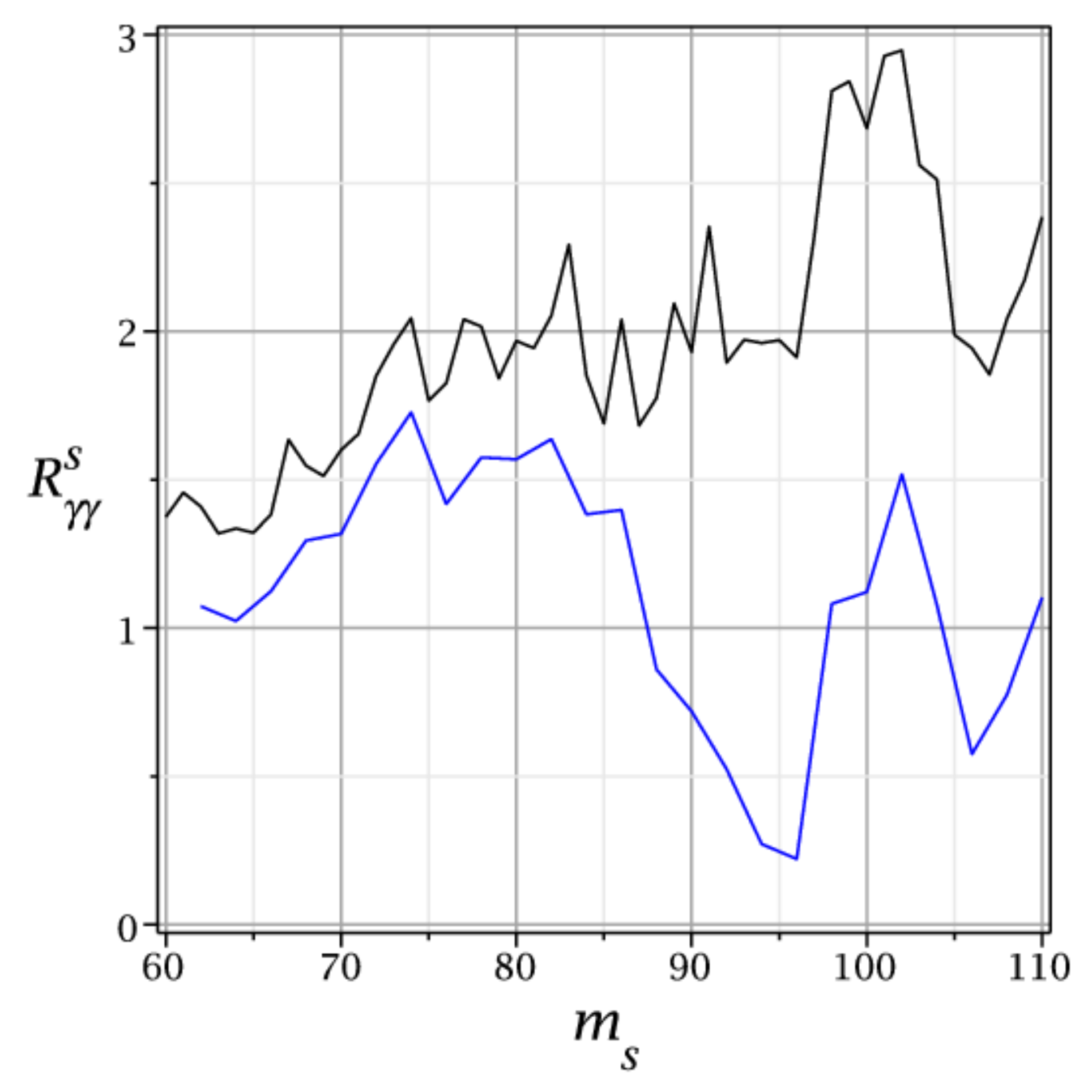}
    \caption{The predictions for $R_{\gamma\gamma}^{(s)}$ assuming maximal value of $\ov{g}_s^2$ consistent with the LEP $s\to jj$ data as a function of $m_s$. The black line corresponds to the case when the $sb\ov{b}$ and $s\tau\ov{\tau}$ couplings are suppressed to zero. The blue line correspond to the minimal suppression of the  $sb\ov{b}$ and $s\tau\ov{\tau}$ couplings required to satisfy the LEP constraints on $\xi_{b\ov{b}}^2$.
    }
    \label{fig:Rgam_s}
  \end{center}
\end{figure}

It should be clear from the previous section that the scenario with a strong suppression of the $sb\ov{b}$ and $s\tau\ov{\tau}$ couplings can be constrained by the precision measurements of the 125 GeV Higgs couplings. Even more interesting is the fact that the LHC is already well prepared for a discovery of $s$. This is because in this scenario the total decay width of $s$ is strongly reduced so all the $s$ branching ratios, except those for the $s$ decays to the down-type fermions, are strongly enhanced.

Particularly interesting is the $\gamma\gamma$ final state.
\footnote{The possibility of large $\gamma\gamma$ rate enhancement for the singlet-like NMSSM boson was noticed in Ref.~\cite{Ellwanger_Rgam_s}. In contrast to the present paper, in Ref.~\cite{Ellwanger_Rgam_s} small values of $\tan\beta\approx3$ were considered. }
 In Figure \ref{fig:Rgam_s} we present the predictions for $R_{\gamma\gamma}^{(s)}$ assuming maximal value of $\ov{g}_s^2$ consistent with the LEP $s\to jj$ data (corresponding to maximal value of $\Delta_{\rm mix}$ allowed by the LEP data) as a function of $m_s$. In the extreme case when the $sb\ov{b}$ and $s\tau\ov{\tau}$ couplings are suppressed to zero (the black line in Figure \ref{fig:Rgam_s}), the $\gamma\gamma$ signal from $s$ decays is stronger than that from the SM Higgs with the same mass for the whole range of $m_s$. For $m_s$ around 100 GeV the enhancement can almost reach a factor of three. 

As we already mentioned, it is not necessary to suppress the $sb\ov{b}$ and $s\tau\ov{\tau}$ couplings exactly to zero. In fact, it is enough to suppress them to the level for which the LEP constraints on $\xi_{b\ov{b}}^2$ are satisfied. The blue line in Figure \ref{fig:Rgam_s} correspond to the minimal suppression of $sb\ov{b}$ and $s\tau\ov{\tau}$ couplings required to satisfy the constraints on $\xi_{b\ov{b}}^2$. Even in this case $R_{\gamma\gamma}^{(s)}>1$ for a wide range of $m_s$ between about 60 and 90 GeV, and around 100 GeV. Small values of $R_{\gamma\gamma}^{(s)}$ are possible only in the few-GeV interval around 95 GeV where the LEP limits on $\xi_{b\ov{b}}^2$ and $\xi_{jj}^2$ are comparable.

Despite such significant enhancement of the $\gamma\gamma$ rate the LHC experiments do not constrain this scenario for $m_s$ below 110 GeV because the Higgs data have not been analysed in that region. For $m_s=110$ GeV, the current observed CMS upper limit \cite{CMS_gamgam} (based on 5 and 20 fb$^{-1}$ of the LHC data at 7 and 8 TeV, respectively) on $R_{\gamma\gamma}^{(s)}$ is about 0.6 which already constrain the allowed values of $\ov{g}_s^2$, thus also $\Delta_{\rm mix}^{\rm max}$, for this particular mass. Therefore, one can expect that the LHC searches are sensitive enough to probe this scenario in the $\gamma\gamma$ channel also for smaller values of $m_s$.

Since the expected limit on $R_{\gamma\gamma}^{(s)}$ for $m_s=110$ GeV with the current data is about 0.6 \cite{CMS_gamgam}, and the sensitivity gets worse quite slowly when the mass goes down,  a naive extrapolation of the available analyses suggests that the LHC could have already set the limits on $R_{\gamma\gamma}^{(s)}\sim\mathcal{O}(1)$ for masses below $100$ GeV using the available data if these were analysed.

\section{Conclusions}
\label{sec6}

We have studied in detail the mixing between the three physical scalars $s$, $h$ and $H$ of the CP-even scalar sector of the NMSSM. In a large parameter range,  it can
lead to several interesting, often correlated, effects. First of all, the $\hat{s}-\hat{h}$ mixing can give 6-8 GeV  contribution to the mass of the SM-like scalar $h$ in the moderate and large
$\tan\beta$ region and with $\lambda\sim\mathcal{O}(0.1)$. This is interesting because the 125  GeV  mass is then obtained with significantly lower stop masses in the stop-top loops. The geometric mean of the stop masses, $M_{\rm SUSY}$, can be below about 400 GeV (2 TeV) for the maximal contribution from stop mixing (with no stop mixing at all).
Thus, the NMSSM is interesting also beyond the usually considered region of low $\tan\beta$ and $\lambda\sim\mathcal{O}(1)$.

The $\hat{s}-\hat{h}$ mixing contribution to $m_h$ depends mainly on the mixing angle between the two fields (i.e.\ on the $sZZ$ coupling $\ov{g}_s$)  and on their mass difference.  Thus the effect is
constrained by the LEP bounds on $\ov{g}_s$ vs $m_s$ obtained  assuming for $s$ the SM Higgs branching ratios and with $b$ and $\tau$ identification in  the final state, or without such particle
identification assuming BR$(s\to{\rm hadrons})=1$. The two experimental bounds almost overlap for $m_s$ in the 5 GeV-interval around $95$ GeV but for other $s$ masses the bound based on the detection of two non-identified hadronic jets is much weaker. For $m_s$ in the 5 GeV-interval around $95$ GeV   the $\mathcal{O}(6\ {\rm GeV})$ mixing contribution to $m_h$  can thus be obtained independently of the decay modes of $s$.
However, for other values of $m_s$ the mixing contribution is much smaller, unless the LEP bound based on the $b$ and $\tau$ identification is evaded, i.e. if  $s\to b\bar{b}$ is suppressed strongly enough.
In the latter case, the 5-8 GeV effect is obtained for the range 60-110 GeV of $m_s$, consistently with the LEP bound based on the search for two hadronic jets, with BR$(s\to{\rm hadrons})=1$.

Interestingly enough, a strong $s\to b\bar{b}$ suppression can be present due to the $\hat{s}-\hat{H}$ mixing (with negligible effect on the $\hat{s}-\hat{h}$ sector),  which is important in the considered  region because of the
$\tan\beta$ enhancement of the scalar down quark couplings. Thus the LEP bounds can be evaded. The lightest scalar $s$ has then enhanced branching ratios into $ZZ^*$, $WW^*$ and $\gamma \gamma$. The latter
one is a particularly promising signature for the LHC searches for a scalar lighter than 110 GeV, with suppressed $b\bar{b}$ decay channel. The signal strength in the $\gamma\gamma$ channel of this scalar may be larger than that of the SM Higgs, even by a factor of three. In fact, if such singlet-like scalar with mass below 110 GeV really exists it could have already been discovered at the LHC if the already collected data were analysed in this range of masses. Thus, we strongly encourage the ATLAS and CMS collaborations to extend their Higgs searches in the $\gamma\gamma$ channel to masses in the 60-110 GeV range.

The $\hat{s}-\hat{H}$ mixing modifies also the $h$ decays, in a way anti-correlated with the $s$ decays.  The ones suppressed for  $s$ are enhanced for $h$ and vice versa. Thus, the large
mixing contribution to $m_h$ can be present together with a variety  of interesting patterns for the $h$ production and decays. If $m_s$ is between about 90 and 105 GeV, the mixing correction to $m_h$ exceeding 5 GeV does not require
the suppression of the $sb\bar{b}$ coupling and e.g. BR$(h\to \gamma\gamma)$ can be either enhanced or suppressed as compared to the SM prediction. If $m_s$  is smaller or larger than the values given above, the
large mixing effect is generically correlated with suppressed rates in $ZZ$, $WW$ and $\gamma\gamma$ channels and enhanced ones in $bb$, $\tau\tau$ channels for $h$. The magnitude of that suppression (enhancement)
depends  on the particular choice of parameters. 

The effects considered in this paper do not require any particular fine tuning of the NMSSM parameters  and are present in a large part of parameter space.

\section*{Acknowledgments}

MO and SP have been supported by National Science Centre under research grants DEC-2011/01/M/ST2/02466, DEC-2012/04/A/ST2/00099, DEC-2012/05/B/ST2/02597.
MB has been partially supported by the Foundation for Polish Science through its programmes HOMING PLUS and START. MB would like to thank the CERN theory division for hospitality during the final stage of this work. The visit of MB to CERN was possible thanks to the National Science Centre research grant DEC-2011/01/M/ST2/02466.


\end{document}